%% file: main.tex
\documentclass[runningheads]{llncs}
\usepackage[T1]{fontenc}
\usepackage{graphicx,verbatim}
\usepackage[T1]{fontenc}
\usepackage{graphicx}
\usepackage{diagbox}
\usepackage{amsfonts}
\usepackage{multirow}
\usepackage{enumitem}
\usepackage{amsmath}

\usepackage{hyperref}
\usepackage[capitalise]{cleveref}
\usepackage{xcolor,soul}
\hypersetup{
    colorlinks=true,
    linkcolor=blue,
    filecolor=cyan,      
    urlcolor=magenta,
    }
\usepackage{booktabs}
\usepackage{color}

\usepackage{bbm}
\usepackage{float}
\usepackage{color, colortbl}
\usepackage{amssymb}  
\usepackage{amsmath}
\usepackage{pifont}   
\usepackage{subcaption} 
\usepackage{tikz}  
\begin{document}
\title{Spatial Masked-Set Learning for Sparse Multi-Shell Diffusion MRI Signal Synthesis}
\titlerunning{Multi-Shell dMRI Signal Synthesis}
%
\author{Yousef Sadegheih\inst{1}\orcidID{0009-0003-1766-5519} \and
Pratibha Kumari\inst{1,2}\orcidID{0000-0003-3681-3700} \and
Dorit Merhof\inst{1,3}\orcidID{0000-0002-1672-2185}}
\authorrunning{Y. Sadegheih et al.}
%
\institute{Faculty of Informatics and Data Science, University of Regensburg, Regensburg, 93053, Germany \and
Department of Data Science and Engineering, IISER Bhopal, India \and
Fraunhofer Institute for Digital Medicine MEVIS, Bremen, Germany \\
\email{dorit.merhof@ur.de}}


  
\maketitle              
\begin{abstract}
Dense multi-shell diffusion MRI provides rich q-space information but requires long acquisition times. We propose a spatial masked-set framework for sparse multi-shell diffusion MRI signal synthesis. The model treats observed measurements as an unordered set, uses a local $3 \times 3 \times 3$ neighborhood for spatial context, and predicts radial-order-6 SHORE coefficients for the center voxel. The coefficients can then be decoded analytically to synthesize signals at arbitrary q-space locations. Training combines shell-wise gradient dropping, dense signal supervision, and rotation-consistent SHORE targets so that sparse input signals remain aligned with their coefficient supervision under augmentation. We evaluate on held-out HCP100 white-matter voxels by retaining limited subsets of measured diffusion-weighted signals from the reference acquisition. The proposed method achieves lower signal NMSE than both analytical q-space models and a state-of-the-art continuous dMRI signal synthesis model designed for arbitrary input and output q-space sampling. In the $b=1000$ setting with 10 input gradients, it achieves $2.70\%$ NMSE, a $22.4\%$ relative reduction over this continuous model. Fractional anisotropy on reconstructed $b=1000$ signals provides a complementary tensor-derived endpoint, with analytical models remaining competitive for FA despite higher dense-signal NMSE across the evaluated q-space. The implementation is available on \href{https://github.com/xmindflow/SHOREPred}{https://github.com/xmindflow/SHOREPred}.

\end{abstract}
\input{introduction}
\input{method}
\input{experiment}
\input{results}
\input{ablations_and_limitations}
    

\begin{credits}
\subsubsection{\ackname} This work was supported by the German Research Foundation (Deutsche Forschungsgemeinschaft, DFG) under the grant no. 417063796. Further, the authors gratefully acknowledge the computational and data resources provided by the Leibniz Supercomputing Center (\href{https://www.lrz.de}{www.lrz.de}).
\subsubsection*{Ethics Statement.}
This study uses publicly available, de-identified Human Connectome Project data; no new human subject data were collected for this work.

\subsubsection{\discintname}
The authors have no competing interests to declare that are
relevant to the content of this article.
\end{credits}

%
%
%
\bibliographystyle{splncs04}
\bibliography{ref}
%




\end{document}

%% file: introduction.tex
\section{Introduction}

Diffusion magnetic resonance imaging (dMRI) characterizes tissue microstructure by measuring signal attenuation across gradient directions and b-values. Dense multi-shell acquisitions support rich q-space modeling but increase scan time, motion sensitivity, and the burden of large studies and clinical protocols \cite{van2013wu,karimi2024diffusion}. Sparse reconstruction therefore aims to recover dense diffusion information from fewer measurements without committing to a single fixed acquisition design.

Learning-based methods have progressed from predicting diffusion-derived quantities or missing shells \cite{golkov2016q,koppers2016diffusion,gibbons2019simultaneous} to encoder-decoder, residual, attention, and shell-specific synthesis models \cite{jha2020multi,dugan2023multi}. Many such methods still assume a fixed number and ordering of input measurements or a specific source-target shell configuration. Implicit neural representations (INRs) and related continuous models provide a natural way to represent signals over coordinates \cite{molaei2023implicit,hendriks2023nesh,wu2024csr,wu2025sarl}, but existing dMRI synthesis settings often focus on a particular shell, a specific source-to-target mapping, or a fixed output sampling pattern, which can require separate models or retraining for different acquisition scenarios. DISCUS addresses arbitrary input and output q-space samplings with geometric deep learning \cite{ewert2024discus}. Analytical q-space models such as Simple Harmonic Oscillator based Reconstruction and Estimation (SHORE) and Mean Apparent Propagator MRI (MAP-MRI) also synthesize signals at arbitrary q-space locations through continuous bases \cite{ozarslan2009simple,ozarslan2013mean,merlet2013continuous}, but their fits become unstable when very few angular and radial samples are observed. We study a complementary route for this low-gradient regime: predicting a continuous SHORE representation from arbitrary observed multi-shell measurements, while using local spatial context and dense signal supervision to stabilize sparse signal synthesis.

We propose a spatial masked-set framework for this setting. The model receives sparse diffusion measurements from a local $3 \times 3 \times 3$ neighborhood, represents each observation by normalized q-space coordinates, b-value, and signal intensity, and predicts radial-order-6 SHORE coefficients for the center voxel. SHORE coefficients act as a compact bridge between arbitrary sparse inputs and continuous dense signal synthesis: once predicted, they are decoded with the SHORE basis at arbitrary q-space locations. Spatial context is injected before mask-aware permutation-invariant pooling over the observed measurements, so the network can use local coherence without requiring a fixed q-space ordering.

Training uses sparse inputs with dense supervision. Shell-wise gradient dropping simulates reduced protocols, while the model is supervised in SHORE-coefficient space and through reconstruction of the full diffusion signal. To keep augmentation physically consistent, rotation-consistent SHORE targets are recomputed after random gradient rotations before coefficient standardization. Our contributions are: \ding{182} masked-set SHORE coefficient prediction for arbitrary sparse multi-shell inputs; \ding{183} a compact spatial architecture using $3 \times 3 \times 3$ context before q-space pooling; \ding{184} sparse-input, dense-output training with rotation-consistent SHORE augmentation and supervision; and \ding{185} an evaluation on the 100-subject Human Connectome Project subset (HCP100) showing lower signal NMSE than a state-of-the-art continuous dMRI reconstruction model and analytical q-space models across the evaluated sparse single- and multi-shell signal-generation settings.

%% file: method.tex
\section{Method}

\subsection{Problem Formulation and SHORE Representation}
\label{subsec:problem_formulation}

Let $v_c$ be the center voxel and let $\mathcal{P}(v_c)$ be its local $3 \times 3 \times 3$ neighborhood. The full acquisition contains $G$ diffusion encodings with $b$-values $b_g$ and unit gradient directions $\mathbf{g}_g=(g_{x,g},g_{y,g},g_{z,g})^\top$. Sparse acquisition is represented by a mask $\mathbf{m}\in\{0,1\}^G$, with observed set $\mathcal{O}=\{g:m_g=1\}$. For each voxel $u\in\mathcal{P}(v_c)$ and encoding $g$, we use the descriptor
\begin{equation}
\mathbf{x}_{u,g}=
\left[\sqrt{\tilde b_g}g_{x,g},\sqrt{\tilde b_g}g_{y,g},\sqrt{\tilde b_g}g_{z,g},\tilde b_g,s_{u,g}\right]^\top,\qquad
\tilde b_g=b_g/b_{\max},
\label{eq:input_descriptor}
\end{equation}
where $s_{u,g}$ is normalized by the mean $b=0$ signal. The first three entries encode normalized q-space position, the fourth preserves shell information, and the fifth is the measured signal. The input is therefore the masked set $\mathbf{X}=\{\mathbf{x}_{u,g}:u\in\mathcal{P}(v_c),g=1,\ldots,G\}$ together with $\mathbf{m}$.

The network predicts center-voxel SHORE coefficients instead of signal intensities tied to a prescribed output gradient table. For radial order 6, the coefficient vector $\mathbf{c}_{v_c}\in\mathbb{R}^{50}$ defines a continuous signal through $\mathbf{s}(Q)=\Phi(Q)\mathbf{c}_{v_c}$, where $\Phi(Q)$ is the SHORE basis evaluated at query q-space locations $Q$ \cite{ozarslan2009simple,merlet2013continuous}. This makes the prediction independent of a particular output gradient table: once coefficients are estimated, signals can be synthesized analytically at arbitrary directions and b-values.

\begin{figure}
    \centering
    \includegraphics[width=\linewidth]{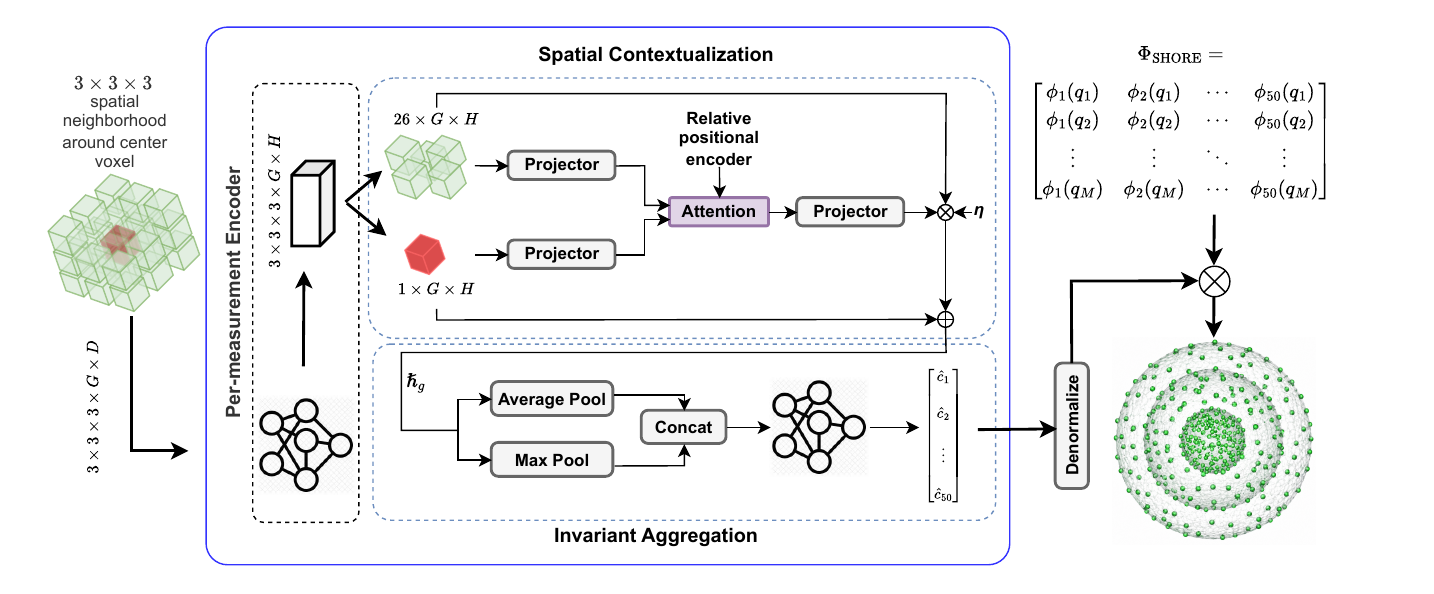}
    \caption{Overview of the spatial masked-set SHORE predictor. Per-encoding features from a $3 \times 3 \times 3$ neighborhood are spatially contextualized, pooled over the observed q-space set, and mapped to 50 standardized SHORE coefficients, which are denormalized and decoded with the SHORE basis to synthesize dense diffusion signals.}
    \label{fig:model_overview}
\end{figure}

\subsection{Spatial Masked-Set Network}
\label{subsec:network}

Fig.~\ref{fig:model_overview} summarizes the proposed network. The design follows two constraints imposed by sparse dMRI: the observed gradients may vary across acquisitions, and their input order is arbitrary. The model is therefore permutation-invariant over observed q-space samples while still using their physical coordinates through Eq.~\eqref{eq:input_descriptor}, following the set-function principle of shared encoding and symmetric pooling \cite{zaheer2017deep,lee2019set}. Unobserved measurements are excluded by the mask during encoding, spatial attention, and q-space pooling, so gradient dropping is treated as missing measurements rather than as zero-valued signal entries.

Each descriptor is encoded by a shared MLP $\phi$ as $\mathbf{h}_{u,g}=\phi(\mathbf{x}_{u,g})\in\mathbb{R}^{H}$ with $H=160$. The encoder has hidden widths 80 and 160 with layer normalization, GELU activations, and dropout. Before pooling over q-space, the center feature for each observed encoding is refined using neighboring voxels measured at the same encoding. For each non-center neighbor $n$, let $\mathbf{r}_n$ be its voxel-grid offset from the center, $\bar{\mathbf{r}}_n$ the component-wise offset normalized by the half-window radius, and $\mathbf{p}_n=[\mathbf{r}_n;\bar{\mathbf{r}}_n;|\mathbf{r}_n|;\|\mathbf{r}_n\|_2^2]\in\mathbb{R}^{10}$ the relative-position descriptor. Spatial attention is
\begin{equation}
\begin{aligned}
e_{n,g}&=\mathbf{w}^{\top}\tanh(W_q\mathbf{h}_{c,g}+W_k\mathbf{h}_{n,g}+W_p\mathbf{p}_n),\\
\alpha_{n,g}&=\operatorname{softmax}_{n\in\mathcal{N}^{\mathrm{valid}}_{c,g}}(e_{n,g}),\\
\bar{\mathbf{h}}_g&=\mathbf{h}_{c,g}+\boldsymbol{\eta}\odot
\sum_{n\in\mathcal{N}^{\mathrm{valid}}_{c,g}}\alpha_{n,g}\mathbf{h}_{n,g}.
\end{aligned}
\label{eq:spatial_context}
\end{equation}
Here $\mathcal{N}^{\mathrm{valid}}_{c,g}$ contains neighbors for which measurement $g$ is available, the attention dimension is 80, and the learnable context scale $\boldsymbol{\eta}\in\mathbb{R}^{H}$ is initialized to $10^{-3}$. This residual scaling lets training start from the center-voxel signal and learn how much local anatomical context to use.

The contextualized center features are aggregated only over observed measurements using mask-aware mean-max pooling,
\begin{equation}
\mathbf{z}=
\left[
|\mathcal{O}|^{-1}\sum_{g\in\mathcal{O}}\bar{\mathbf{h}}_g
;\;
\max_{g\in\mathcal{O}}\bar{\mathbf{h}}_g
\right],
\qquad
\hat{\mathbf{c}}^{\mathrm{std}}=\rho(\mathbf{z}),
\label{eq:coefficient_prediction}
\end{equation}
where the maximum is element-wise. Mean pooling captures the overall sparse q-space trend, max pooling preserves salient directional evidence, and the prediction head $\rho$ maps the $2H$ pooled representation through hidden widths 320 and 160 to the 50 standardized SHORE coefficients.

\subsection{Sparse Input, Dense Supervision, and SHORE Normalization}
\label{subsec:sparse_dense_supervision}

Training simulates reduced acquisitions by applying shell-wise masks to the full diffusion signal. For each selected shell, a fixed number of directions is retained from antipodal subsets to preserve approximate angular coverage; the remaining measurements are hidden from the model. The target remains dense: predicted coefficients are supervised both as SHORE coefficients and through reconstruction of the full diffusion signal. This sparse-input, dense-output objective matches the intended inference setting, where only sparse measurements are acquired but a dense continuous q-space representation is desired.

SHORE coefficients are indexed by radial order $n$, angular degree $l$, and angular order $m$. Coefficients with the same $(n,l)$ form a rotational block $\mathcal{B}_{n,l}=\{c_{n,l,m}:m=-l,\ldots,l\}$. We normalize these blocks in a way that respects this structure. The rotationally invariant $l=0$ coefficient for each $n$ is centered and scaled using training-set statistics $\mu_{n,0}$ and $a_{n,0}$. For $l>0$, a spatial rotation does not simply permute angular coefficients; it applies a linear transform within the $m=-l,\ldots,l$ components of each $(n,l)$ block. We therefore avoid component-wise centering, which would introduce orientation-dependent offsets into a subspace that should rotate as a block. Instead, each block is normalized only by a shared root-mean-square scale, preserving the relative angular geometry while balancing coefficient magnitudes numerically:
\begin{equation}
a_{n,l}=\max\!\left\{\sqrt{\frac{1}{N(2l+1)}\sum_{i=1}^{N}\sum_{m=-l}^{l}\left(c_{n,l,m}^{(i)}\right)^2},\epsilon\right\},\quad l>0.
\label{eq:lgt0_scale}
\end{equation}
Here $\epsilon>0$ defines the minimum scale for each block, preventing division by zero during normalization. Standardization is therefore $c_{n,0,0}^{\mathrm{std}}=(c_{n,0,0}-\mu_{n,0})/a_{n,0}$ for $l=0$, and $c_{n,l,m}^{\mathrm{std}}=c_{n,l,m}/a_{n,l}$ for $l>0$. All statistics are computed on the training set and fixed for validation and testing.

The coefficient loss is computed in standardized space, $\mathcal{L}_{\mathrm{coeff}}=C^{-1}\|\hat{\mathbf{c}}^{\mathrm{std}}-\mathbf{c}^{\mathrm{std}}\|_2^2$. For signal supervision, predictions are denormalized back to SHORE coefficient scale, decoded as $\hat{\mathbf{s}}=\Phi\hat{\mathbf{c}}$, and compared with the full normalized signal using $\mathcal{L}_{\mathrm{signal}}=G^{-1}\|\hat{\mathbf{s}}-\mathbf{s}\|_2^2$. The total objective is
\begin{equation}
\mathcal{L}=\lambda_c\mathcal{L}_{\mathrm{coeff}}+\lambda_s\mathcal{L}_{\mathrm{signal}},
\qquad \lambda_c=1,\quad \lambda_s=10.
\label{eq:total_loss}
\end{equation}
The stronger signal-domain weight encourages accurate dense reconstruction while the coefficient term keeps the learned representation close to the SHORE target space.

\subsection{Rotation-Consistent SHORE Supervision}
\label{subsec:rotation_consistent_supervision}

Because q-space coordinates are explicit inputs, training only on one gradient-table orientation can encourage orientation-specific shortcuts. We therefore apply random rotations during training. For $R\in SO(3)$, gradients are transformed as $\mathbf{g}'_g=R\mathbf{g}_g$; $b$-values and signal intensities are unchanged, but the descriptor in Eq.~\eqref{eq:input_descriptor} is recomputed with the rotated directions.

The SHORE target must rotate with the input geometry. Reusing the original coefficient vector would be inconsistent because SHORE coefficients are basis-dependent. Let $\Phi$ and $\mathbf{c}$ be the original SHORE basis and coefficient vector, and let $\mathbf{s}_{\mathrm{ref}}=\Phi\mathbf{c}$ be the reference dense signal. After rotating the gradient table, we construct the rotated basis $\Phi'$ and solve
\begin{equation}
\mathbf{c}'=\arg\min_{\mathbf{u}\in\mathbb{R}^{C}}\|\Phi'\mathbf{u}-\mathbf{s}_{\mathrm{ref}}\|_2^2,
\qquad
\left({\Phi'}^{\top}\Phi'\right)\mathbf{c}'={\Phi'}^{\top}\mathbf{s}_{\mathrm{ref}}.
\label{eq:rotated_coefficients}
\end{equation}
In practice, we solve these normal equations with a Cholesky factorization of ${\Phi'}^{\top}\Phi'$. The rotated coefficients are then standardized with the blockwise SHORE normalization defined in Sec.~\ref{subsec:sparse_dense_supervision}. This keeps the augmented input and supervision target in the same rotated acquisition geometry while preserving a stable coefficient distribution. The procedure is important for $l>0$ blocks, where rotations linearly transform angular orders within each $(n,l)$ block; it is exactly this coupling that motivates the shared block scaling used for SHORE normalization.

%% file: experiment.tex
\section{Experiments}
\label{sec:experiments}

\subsection{Dataset and Preprocessing}
\label{subsec:dataset_preprocessing}

We evaluate on the HCP100 subset of the Human Connectome Project \cite{van2013wu,sotiropoulos2013advances}. Each subject has three nonzero shells at approximately $b=1000$, $2000$, and $3000~\mathrm{s/mm^2}$, with 90 diffusion-weighted measurements per shell and 18 $b=0$ volumes. The acquisition uses a pulsed-gradient spin-echo sequence with $\Delta\approx43.1~\mathrm{ms}$ and $\delta\approx10.6~\mathrm{ms}$, which are needed by analytical q-space models such as SHORE and MAP-MRI. Subjects are split at the subject level into 80 training, 10 validation, and 10 test subjects.

Training and evaluation are performed in white matter masks obtained with FSL \cite{jenkinson2012fsl}. We focus on white matter because sparse angular reconstruction is most relevant where diffusion has strong directional structure and multi-shell q-space information is expected to matter. Including gray matter, CSF, or partial-volume tissue would add many low-angular-contrast voxels and could obscure differences between reconstruction methods. Signals are normalized voxel-wise by the mean $b=0$ signal, $s_g=S_g/\overline{S}_0$, to reduce global intensity differences across voxels and subjects.

Ground-truth SHORE coefficients are fitted voxel-wise from the full reference acquisition using radial order 6 and the Diffusion Imaging in Python (DIPY) SHORE implementation \cite{garyfallidis2014dipy,ozarslan2009simple}. The resulting 50-dimensional coefficient vectors provide dense supervision targets, and precomputed SHORE basis matrices decode predicted coefficients back to diffusion signals.

\subsection{Sparse Sampling Protocol}
\label{subsec:sparse_sampling_protocol}

Sparse acquisitions are simulated by masking diffusion measurements shell-wise. For each included shell, we retain $k\in\{5,10,15,20,25,30,40,50\}$ directions and hide the rest. In multi-shell settings with $S$ shells, this gives $kS$ observed diffusion-weighted measurements. Retained directions are selected from antipodal groups with approximately balanced angular coverage, and the same sparse masks are used for all methods to ensure a matched comparison across extremely sparse and moderately sparse regimes. The proposed model is trained once and evaluated across all shell combinations and sampling budgets without configuration-specific retraining.

\subsection{Comparison Methods}
\label{subsec:comparison_methods}

We compare against DISCUS, a recent geometric deep learning method for continuous dMRI reconstruction with arbitrary input and output samplings \cite{ewert2024discus}. We trained DISCUS from its public source code using the hyperparameter settings provided by the authors in the code release and paper. We also evaluate analytical SHORE and MAP-MRI models using DIPY \cite{garyfallidis2014dipy}. SHORE uses radial order 6 to match our target representation \cite{ozarslan2009simple}; MAP-MRI uses radial order 4, following the Sparse Reconstruction Challenge setting \cite{ozarslan2013mean,ning2015sparse}. Analytical models are fitted only from the sparse observed measurements and then evaluated on the full reference signal.

\subsection{Evaluation Metrics}
\label{subsec:evaluation_metrics}

Signal fidelity is measured by patient-macro normalized mean squared error (NMSE). Let $\mathcal{T}$ be the test subjects, $\Omega_p$ the white matter voxels for subject $p$, and $\mathcal{G}$ the dense evaluation gradient table. We compute
\begin{equation}
\mathrm{NMSE}=
\frac{1}{|\mathcal{T}|}\sum_{p\in\mathcal{T}}
\frac{1}{|\Omega_p|}\sum_{v\in\Omega_p}
\frac{\sum_{g\in\mathcal{G}}(\hat{s}_{p,v,g}-s_{p,v,g})^2}
{\sum_{g\in\mathcal{G}}s_{p,v,g}^2+\varepsilon},
\label{eq:nmse_metric}
\end{equation}
where $\varepsilon$ prevents division by zero. This averages voxel errors within each subject and then averages subjects equally, rather than pooling all test voxels. To assess whether synthesized $b=1000$ signals preserve tensor-derived information, we also report patient-macro fractional anisotropy error,
\begin{equation}
\mathrm{MSE}_{\mathrm{FA}}=
\frac{1}{|\mathcal{T}|}\sum_{p\in\mathcal{T}}
\frac{1}{|\Omega_p|}\sum_{v\in\Omega_p}
(\widehat{\mathrm{FA}}_{p,v}-\mathrm{FA}_{p,v})^2.
\label{eq:fa_metric}
\end{equation}
The reference and reconstructed FA values are obtained by fitting diffusion tensors to the corresponding $b=1000$ signals using the same gradient table. NMSE and $\mathrm{MSE}_{\mathrm{FA}}$ are reported as percentages.

\subsection{Implementation Details}
\label{subsec:implementation_details}

The model is implemented in PyTorch 2.5.0 \cite{paszke2019pytorch}; SHORE, MAP-MRI, and tensor fitting use DIPY 1.9.0 \cite{garyfallidis2014dipy}. Training uses Adam \cite{kingma2014adam} with learning rate $10^{-3}$, weight decay $10^{-4}$, a polynomial learning-rate scheduler with power 0.9, and batch size 512. The final model is trained for 250,000 optimization steps on 1 NVIDIA A100 GPU with 40 GB VRAM, requiring approximately 10 GPU-hours. The loss weights are $\lambda_c=1$ and $\lambda_s=10$, and random rotation augmentation is applied to training samples with probability 0.25. The final network contains approximately 0.23 million trainable parameters, comparable to the approximately 0.25 million trainable parameters of DISCUS.

%% file: results.tex
\begin{table*}[!t]
\newcommand{\tabbest}[1]{\textcolor{blue}{#1}}
\newcommand{\tabsecond}[1]{\textcolor{red}{#1}}
\centering
\caption{Dense multi-shell reconstruction from sparse $b=1000$ inputs on the HCP100 test set. The gradient count denotes the number of observed $b=1000$ directions; signal NMSE is computed on the reconstructed $b=1000$, $b=2000$, and $b=3000$ signals with 90 directions per shell, while MSE\textsubscript{FA} is computed from FA maps fitted to the reconstructed $b=1000$ shell. Both metrics are reported as mean (standard deviation) in percentages; \tabbest{blue} and \tabsecond{red} indicate the best and second-best means for each sampling budget.}
\label{tab:b1000_results}
\resizebox{\textwidth}{!}{
\begin{tabular}{l||cc||cc||cc||cc}
\toprule
\multirow{2}{*}{\shortstack{\# input\\directions}} 
& \multicolumn{2}{c||}{DISCUS~\cite{ewert2024discus}} 
& \multicolumn{2}{c||}{SHORE~\cite{ozarslan2009simple}} 
& \multicolumn{2}{c||}{MAP~\cite{ozarslan2013mean}} 
& \multicolumn{2}{c}{Ours} \\ 
\cline{2-9}
& \multicolumn{1}{c|}{MSE\textsubscript{FA} $\downarrow$} & NMSE $\downarrow$ 
& \multicolumn{1}{c|}{MSE\textsubscript{FA} $\downarrow$} & NMSE $\downarrow$ 
& \multicolumn{1}{c|}{MSE\textsubscript{FA} $\downarrow$} & NMSE $\downarrow$ 
& \multicolumn{1}{c|}{MSE\textsubscript{FA} $\downarrow$} & NMSE $\downarrow$ \\ 
\midrule
\multicolumn{1}{c||}{5}  
& \multicolumn{1}{c|}{2.15 (0.043)} & \tabsecond{4.81 (0.011)} 
& \multicolumn{1}{c|}{\tabbest{1.56 (0.081)}} & 12.70 (0.768)
& \multicolumn{1}{c|}{3.19 (0.153)} & 7.29 (0.214)
& \multicolumn{1}{c|}{\tabsecond{1.80 (0.042)}} & \tabbest{4.08 (0.004)} \\

\multicolumn{1}{c||}{10} 
& \multicolumn{1}{c|}{0.83 (0.038)} & \tabsecond{3.48 (0.022)} 
& \multicolumn{1}{c|}{\tabsecond{0.66 (0.069)}} & 10.80 (0.701)
& \multicolumn{1}{c|}{1.05 (0.051)} & 5.34 (0.215)
& \multicolumn{1}{c|}{\tabbest{0.41 (0.028)}} & \tabbest{2.70 (0.003)} \\

\multicolumn{1}{c||}{15} 
& \multicolumn{1}{c|}{0.61 (0.076)} & \tabsecond{3.18 (0.035)} 
& \multicolumn{1}{c|}{\tabsecond{0.41 (0.046)}} & 10.38 (0.647)
& \multicolumn{1}{c|}{0.54 (0.036)} & 4.89 (0.221)
& \multicolumn{1}{c|}{\tabbest{0.27 (0.013)}} & \tabbest{2.49 (0.002)} \\

\midrule
\multicolumn{1}{c||}{20} 
& \multicolumn{1}{c|}{0.50 (0.087)} & \tabsecond{3.05 (0.037)} 
& \multicolumn{1}{c|}{\tabsecond{0.29 (0.035)}} & 10.11 (0.663)
& \multicolumn{1}{c|}{0.34 (0.033)} & 4.68 (0.199)
& \multicolumn{1}{c|}{\tabbest{0.22 (0.010)}} & \tabbest{2.40 (0.003)} \\

\multicolumn{1}{c||}{25} 
& \multicolumn{1}{c|}{0.43 (0.083)} & \tabsecond{2.97 (0.045)} 
& \multicolumn{1}{c|}{\tabsecond{0.22 (0.028)}} & 9.97 (0.655)
& \multicolumn{1}{c|}{0.25 (0.028)} & 4.57 (0.197)
& \multicolumn{1}{c|}{\tabbest{0.17 (0.010)}} & \tabbest{2.34 (0.003)} \\

\multicolumn{1}{c||}{30} 
& \multicolumn{1}{c|}{0.39 (0.076)} & \tabsecond{2.94 (0.040)} 
& \multicolumn{1}{c|}{\tabbest{0.17 (0.021)}} & 9.82 (0.630)
& \multicolumn{1}{c|}{0.21 (0.022)} & 4.51 (0.186)
& \multicolumn{1}{c|}{\tabbest{0.17 (0.011)}} & \tabbest{2.31 (0.003)} \\

\midrule
\multicolumn{1}{c||}{40} 
& \multicolumn{1}{c|}{0.36 (0.063)} & \tabsecond{2.92 (0.040)} 
& \multicolumn{1}{c|}{\tabbest{0.11 (0.013)}} & 9.65 (0.603)
& \multicolumn{1}{c|}{0.18 (0.015)} & 4.41 (0.182)
& \multicolumn{1}{c|}{\tabsecond{0.15 (0.015)}} & \tabbest{2.27 (0.004)} \\

\multicolumn{1}{c||}{50} 
& \multicolumn{1}{c|}{0.36 (0.043)} & \tabsecond{2.93 (0.039)} 
& \multicolumn{1}{c|}{\tabbest{0.07 (0.008)}} & 9.54 (0.600)
& \multicolumn{1}{c|}{0.20 (0.011)} & 4.36 (0.174)
& \multicolumn{1}{c|}{\tabsecond{0.14 (0.015)}} & \tabbest{2.24 (0.005)} \\

\bottomrule
\end{tabular}
}
\end{table*}

\section{Results and Discussion}
\label{sec:results_discussion}

\begin{figure}
    \centering
    \includegraphics[width=\linewidth]{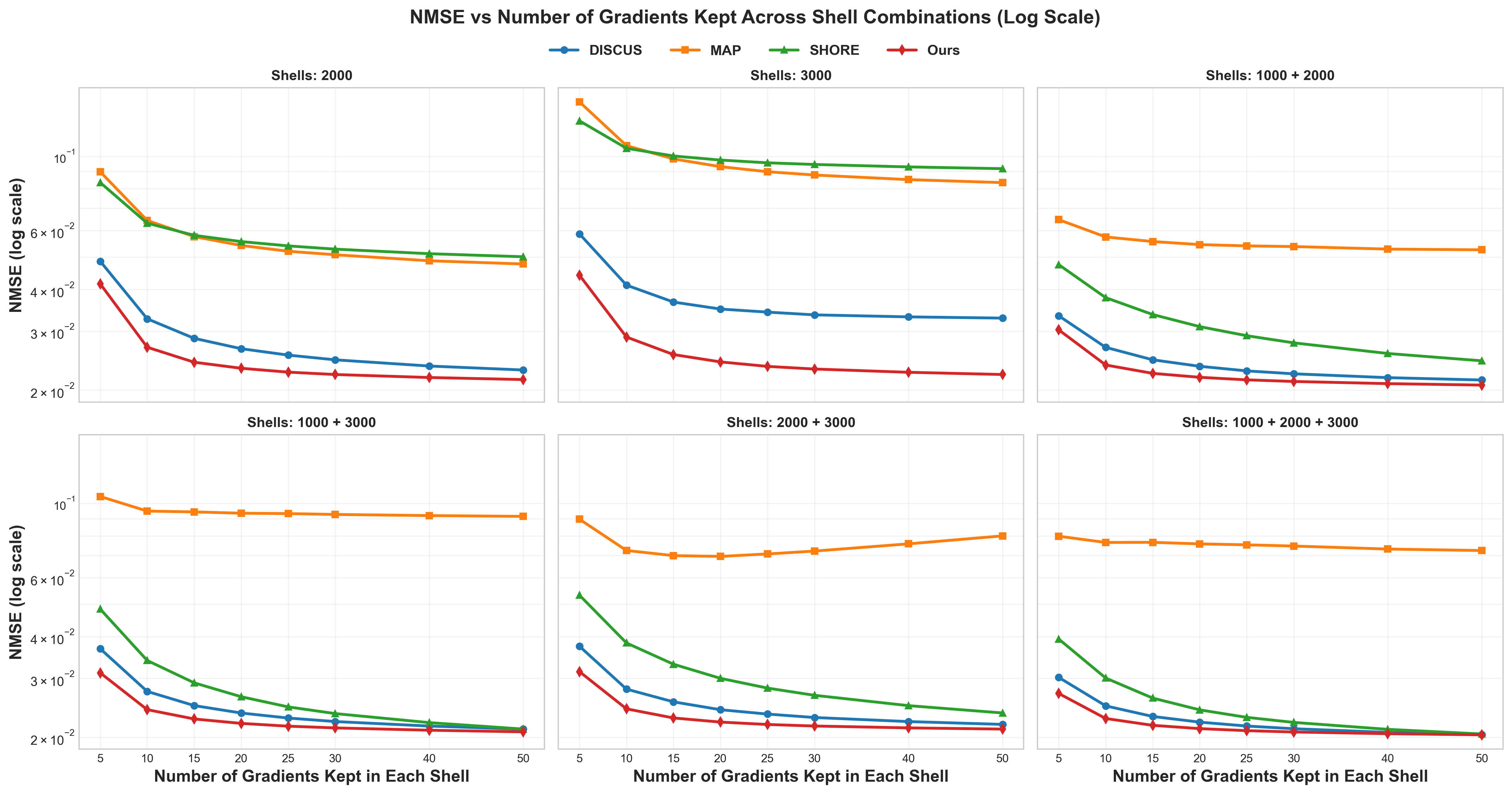}
    \caption{Signal NMSE comparison across sparse single-shell and multi-shell input settings. Each panel specifies the shells available as sparse input; an x-axis value of $k$ means that $k$ directions are retained from each included shell using antipodal sampling. NMSE is computed on dense reconstruction of the $b=1000$, $b=2000$, and $b=3000$ signals with 90 directions per shell. The y-axis is shown on a logarithmic scale. Lower values indicate better reconstruction.}
\label{fig:nmse_shells}
\end{figure}

Table~\ref{tab:b1000_results} and Fig.~\ref{fig:nmse_shells} evaluate the same question at different input sparsity patterns: how well each method recovers the dense multi-shell q-space signal when only a small masked subset is observed. The clearest advantage of the proposed method is in signal NMSE, especially at low gradient counts where analytical basis fitting is underconstrained and the ability to use arbitrary sparse measurements is most important.

For the $b=1000$-only sparse-input protocol, the proposed method gives the lowest dense-signal NMSE at every sampling budget in Table~\ref{tab:b1000_results}. In the most sparse setting, it reduces NMSE by $15.2\%$ relative to DISCUS, $44.0\%$ relative to MAP-MRI, and $67.9\%$ relative to SHORE. The comparison with DISCUS remains favorable as more gradients are retained, with relative NMSE reductions of $22.4\%$ at 10 gradients and $23.5\%$ at 50 gradients. These relative improvements should be interpreted as signal-domain gains under the simulated sparse HCP100 protocol, not as a general claim about all downstream diffusion metrics.

The FA results provide a narrower tensor-derived view of the reconstruction. They test whether the reconstructed $b=1000$ shell preserves anisotropy after diffusion-tensor fitting, whereas signal NMSE evaluates the dense multi-shell signal itself. This explains why the rankings are not identical: the proposed method is best for FA in the intermediate sparse regimes and ties SHORE at 30 retained gradients, while SHORE remains strongest at the sparsest and densest settings. Thus, analytical q-space fitting can preserve a dominant tensor summary well even when its dense multi-shell signal NMSE is substantially worse.

The multi-shell NMSE curves in Fig.~\ref{fig:nmse_shells} are consistent with this signal-domain interpretation. In the low-gradient regime, the proposed method is the strongest method across the plotted shell combinations, with particularly visible separation for higher b-values and multi-shell inputs. DISCUS is usually the closest learning-based comparison method, while SHORE and MAP-MRI improve as more measurements are retained but remain more sensitive to extreme sparsity. At larger sampling budgets, several curves approach each other and ties or near-ties become more common, which is expected because the reconstruction problem becomes less underdetermined and analytical models receive enough measurements for more stable fitting.

Overall, the results support the intended operating point of the method: sparse multi-shell signal synthesis from limited q-space samples in white matter. Under this protocol, masked-set processing with local spatial context improves dense signal reconstruction over DISCUS and the analytical q-space models, while the FA results show that method rankings can change for specific tensor-derived summaries.

%% file: ablations_and_limitations.tex
\section{Ablation Studies}
\label{sec:ablation}

Table~\ref{tab:ablation} separates model design from sparse-input training. The full-observation setting uses all 270 diffusion-weighted measurements as input and reconstructs the full signal, isolating architecture without sparsity-induced information loss. The sparse setting tests whether gradient dropping during training is needed for the challenging $b=1000$ case with 10 retained gradients. We compare against fixed-order multilayer perceptron (MLP) variants to test sensitivity to arbitrary gradient ordering.

\begin{table}[t]
\centering
\caption{Ablation study. NMSE is reported in percent; lower is better. Full observation uses all 270 diffusion-weighted measurements, while sparse input uses $b=1000$ with 10 retained gradients.}
\label{tab:ablation}
\begin{tabular}{llc}
\toprule
Setting & Variant & NMSE (\%) $\downarrow$ \\
\midrule
\multirow{5}{*}{Full observation}
& Fixed-order MLP & 2.25 \\
& MLP with shuffled gradient order & 7.32 \\
& Basic Deep Set & 2.15 \\
& Basic Deep Set + rotation augmentation & 2.07 \\
& Spatial masked-set model + rotation augmentation & \textbf{1.90} \\
\midrule
\multirow{2}{*}{Sparse input}
& Without gradient dropping during training & 6.65 \\
& With gradient dropping during training & \textbf{2.70} \\
\bottomrule
\end{tabular}
\end{table}

The fixed-order MLP performs reasonably only when the gradient order is fixed (2.25\% NMSE), but degrades to 7.32\% when the input order is shuffled. This confirms that it learns position-specific dependencies rather than an unordered q-space representation, which is unsuitable for arbitrary sparse protocols. A basic Deep Set model reduces NMSE to 2.15\%, and rotation augmentation further improves it to 2.07\%, an approximately 4\% relative reduction, suggesting that exposure to additional q-space orientations and corresponding SHORE coefficient values is beneficial. Adding local spatial context on top of rotation augmentation gives the best full-observation result, 1.90\% NMSE, an 8.2\% relative reduction over the augmented Deep Set and 15.6\% over the fixed-order MLP. The trained spatial model also learned a non-negligible residual context scale: the mean absolute learned $\boldsymbol{\eta}$ was 0.51 after initialization at $10^{-3}$. Together with the NMSE improvement, this suggests that the neighbor-context pathway was used rather than suppressed during training.

Sparse-input training is equally important. Without gradient dropping during training, the model reaches 6.65\% NMSE when tested with only 10 $b=1000$ gradients. Training with gradient dropping reduces this to 2.70\%, a 59.4\% relative error reduction. Thus, the final performance comes from the combination of order-invariant q-space processing, rotation-consistent supervision, learned spatial context, and explicit exposure to missing-gradient patterns.

\section{Limitations and Future Work}
\label{sec:limitations}

This study is limited to HCP100 white matter voxels and retrospective sparse masks derived from dense acquisitions. This design isolates sparse q-space reconstruction in anisotropic tissue, but it does not establish whole-brain performance or robustness under prospectively acquired sparse protocols, where motion, noise, scanner differences, and protocol-specific artifacts may differ. The target representation is also fixed to radial-order-6 SHORE, which provides compact analytic decoding but may limit expressiveness in more complex diffusion settings. Finally, we evaluate signal NMSE and FA error; future work should test downstream orientation estimation, tractography, and microstructural modeling, and should assess generalization across datasets, scanners, and acquisition protocols.

\section{Conclusion}
\label{sec:conclusion}

We presented a compact spatial masked-set model for sparse multi-shell dMRI signal synthesis. A single trained model handles all evaluated sparse shell combinations and sampling budgets using local $3 \times 3 \times 3$ context, predicts radial-order-6 SHORE coefficients, and decodes them at arbitrary q-space query locations without retraining for each input mask or output gradient table. On HCP100 white matter, it improves signal NMSE over a state-of-the-art continuous dMRI reconstruction model and analytical q-space models in low-gradient regimes, while remaining competitive for FA reconstruction. The ablations show that permutation-invariant set processing, rotation-consistent supervision, learned spatial context, and sparse-input training each contribute to the final model. These results support spatial masked-set SHORE prediction as an effective modeling strategy for sparse dMRI reconstruction under the evaluated protocol.